\documentclass[runningheads]{llncs}
\usepackage{graphicx}
\usepackage{color}
\usepackage{multirow}
\usepackage{bm}
\usepackage{subcaption}
\usepackage{array}
\usepackage{fontawesome}
\usepackage{url}

\newcolumntype{U}{>{\centering\arraybackslash}p{0.45cm}}

\begin{document}
\title{A Study of Maintainability in Evolving Open-Source Software}
\titlerunning{A Study of Maintainability in Evolving Open-Source Software}
%
\author{Arthur-Jozsef Molnar\orcidID{0000-0002-4113-2953} \and Simona Motogna\orcidID{0000-0002-8208-6949}}
\institute{Faculty of Mathematics and Computer Science, Babe\c{s}-Bolyai University, Cluj-Napoca, Romania
\email{\{arthur, motogna\}@cs.ubbcluj.ro}\\ \url{http://www.cs.ubbcluj.ro}}

%
%
%
\maketitle              
\begin{abstract}
Our study is focused on an evaluation of the maintainability characteristic in the context of the long-term evolution of open-source software. According to well established software quality models such as the ISO 9126 and the more recent ISO 25010, maintainability remains among key quality characteristics alongside performance, security and reliability. To achieve our objective, we selected three complex, widely used target applications for which access to their entire development history and source code was available. To enable cross-application comparison, we restricted our selection to GUI-driven software developed on the Java platform. We focused our examination on released versions, resulting in 111 software releases included in our case study. These covered more than 10 years of development for each of the applications. For each version, we determined its maintainability using three distinct quantitative models of varying complexity. We examined the relation between software size and maintainability and studied the main drivers of important changes to software maintainability. We contextualized our findings using manual source code examination. We also carried out a finer grained evaluation at package level to determine the distribution of maintainability issues within application source code. Finally, we provided a cross-application analysis in order to identify common as well as application-specific patterns.
\keywords{Software quality, Software metrics, Software maintainability, Software evolution, Maintainability index, SQALE model, Technical debt, open-source}
\end{abstract}

\section{Introduction}
\label{sec:introduction}
Maintenance includes all activities intended to correct faults, update the target system in accordance to new requirements, upgrade system performance and adapt it to new environment conditions. As a consequence, maintenance effort becomes very costly, especially in the case of large-scale complex applications, especially since in many cases they include third party components sensitive to updates.

The causes of high maintenance costs can be tracked to multiple reasons. The first such reason regards the inherent complexity of code comprehension tasks, due to the fact that maintenance teams are different from the development team, causing further delays for understanding source code and locating software defects. Another important issue is that maintenance is approached only during the late stages of the development lifecycle, when issues have already built up in the form of technical debt \cite{29}. These reasons can be overcome by considering maintainability issues earlier in the development process and employing existing tool support that can help identify future maintenance ''hospots'', namely those parts of code that can generate more problems. If we consider agile practices, then integration of maintenance tasks with development processes becomes a necessity. When these issues are not addressed at the right moment they tend to accumulate in the form of technical debt that can later lead to crises during which development is halted until the bulk of the issues are addressed \cite{48}. 

The focus of this study regards the long term assessment of maintainability in large software applications; software evolution plays an important part that is observed through the release of a consistent number of software versions. In many of these applications we find that complexity is increased by functionalities end-users rely on. They are usually implemented as plugins, which can create additional dependencies on the main code base. Our empirical investigation targets open source applications where full access to the source code was available over the entire application life span. This not only allows the usage of quantitative quality models based on software metrics, but also facilitates the manual examination of source code, which can be used to understand the rationale behind observed changes to application architecture and structure. 

Previous studies have identified some of the existing relations between the maintainability characteristic and software metric values \cite{7,8,21,22,33,24}. Our goal is to employ several quantitative models having well studied strengths and weaknesses \cite{44} in order to determine some of the patterns in the evolution of open-source software, to understand the rationale behind important changes to source code, as well as to improve our understanding of the quality models and their applicability.

The present study continues our existing research regarding the maintainability of open-source software \cite{44} and brings the following novel contributions: (a) a longitudinal study of software maintainability that covers the entire development history of three complex, open-source applications; (b) a detailed examination of the relation between maintainability, expressed through several quantitative models and software size measured according to several levels of granularity; (c) an examination of sudden changes to maintainability as well as ''slopes'' - significant modifications that occur over the span of several releases; (d) a finer-grained analysis at package and class level regarding software maintainability and its evolution; (e) an analysis of the maintainability models themselves as applied to real-life open-source software systems.

\section{Software Quality Models}
The importance of software quality continues to pose a key interest in both the academic and industry communities after more than 50 years of research and practice. Furthermore, as the number of complex networked systems and critical infrastructures relying on them is increasing, it is expected to remain an issue of continued interest in software research, development and maintenance. Previous research into software quality resulted in a large number of quality models. Most of them describe a set of essential attributes that attempt to characterize the multiple facets of a software system from an internal (developer-oriented), external (client-oriented) or both perspectives. 

\begin{figure*}
  \includegraphics[width=\linewidth]{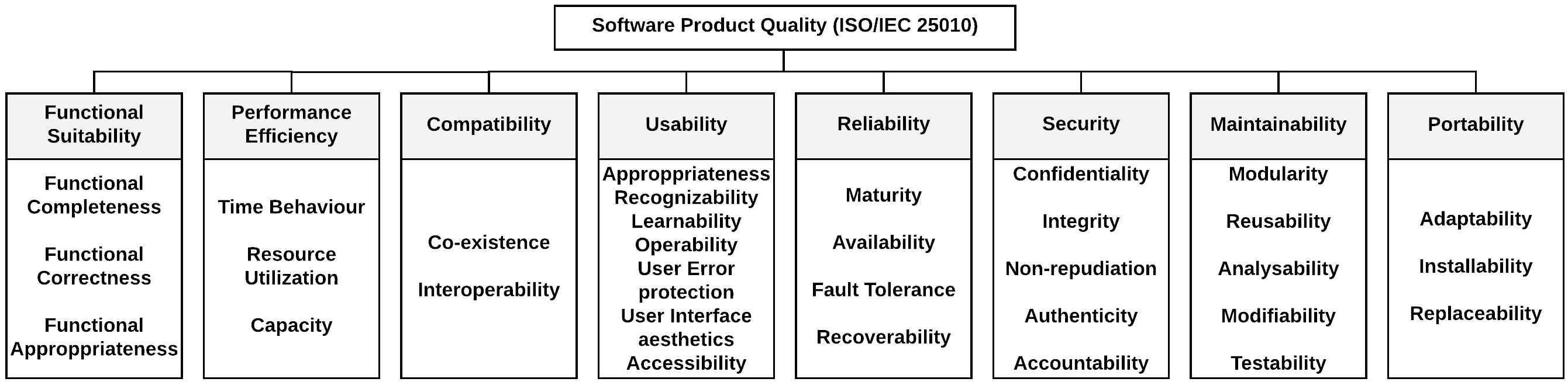}
  \caption{ISO 25010 hierarchical quality model (from \cite{54})}
  \label{fig:ISO}
\end{figure*}

The introduction of the first software quality model is attributed to McCall in 1976, followed by the Dromey model which improved it \cite{36}. Later on, these initial contributions became a part of the ISO 9126 standard, which expressed software quality using a hierarchical model of six characteristics that are comprised of 31 sub-characteristics. The ISO 25010 model \cite{23} illustrated in Figure \ref{fig:ISO} represents the current version, and it considers maintainability as the ensemble of six sub-characteristics: \textit{Modularity}, \textit{Reusability}, \textit{Analysability}, \textit{Modifiability} and \textit{Testability}. Like its previous versions, ISO 25010 does not provide a methodology to evaluate quality characteristics or to improve them, which precludes practitioners from using them directly. However, this shortcoming can be overcome using software metrics, which measure different properties of source code and related artefacts. Basic metrics such as lines of code, number of functions or modules have been widely used and in turn, superseded by the introduction of the object oriented paradigm and its related set of metrics. Nowadays we find a multitude of object oriented metrics \cite{7} defined and used to detect code smells, design flaws or in order to improve maintainability. These metrics were also harnessed by researchers to evaluate software quality in general. However, these tasks have remained difficult and tedious in the context of large-scale software systems.

Authors of \cite{51} study the relation between object-oriented metrics and software defects. They report the response for a class (RFC) and weighted method count (WMC) as most suited for identifying defect potential. A similar study using the Mozilla application suite \cite{52} showed the coupling between objects (CBO) and lines of code (LOC) as accurate fault predictors. These findings were backed by \cite{53}, where a NASA data set was the target of defect estimation efforts, and \cite{49}, where evaluation was carried out using eight C++ applications. This also leads to the issue of the target application's programming language, with authors \cite{22} claiming that metric value expectations have to be adapted to each language in particular. The over-arching conclusion of metric-based evaluations is that further work is required before definitive expectations can be formalized regarding the relation between software quality characteristics and metric values.   

\section{Maintainability Models}
\subsection{Maintainability Index}
\label{sec:MI}
There exists long-term interest regarding the correct estimation of the required effort for maintaining software systems. Initially defined in the late '70, the computational formula for the Maintainability Index (MI) was introduced in 1992 \cite{13}. The formula takes into consideration source code size, measured according to variants of the lines of code metric and two views of complexity expressed in terms of the modular paradigm; they are the number of operations and operators, also known as the Halstead volume and the number of possible execution paths generated by existing conditional and loop statements. The variant employed in our research is \cite{13}:

$MI = 171 - 5.2 * ln(aveV) - 0.23 * aveG - 16.2 * ln(aveSTAT)$

\noindent where \textit{aveV} denotes average Halstead volume, \textit{aveG} is the number of possible execution paths (cyclomatic complexity) and \textit{aveSTAT} is the average number of statements. Several versions of this formula exist, such as considering the LOC metric instead of statement counts, or including the number of lines of comments into the formula. The presented version returns values between 171 (best maintainability) and negative numbers, which are evaluated as very poor maintainability. Several implementations \cite{32} normalize the formula to return values in the $[0,100]$ range by translating all negative values to 0. Different development, metric or code inspection tools compute the MI \cite{14,15,16,32}, and some provide good practices \cite{32}, stating that values below 20 correspond to poor maintainability. 

Several criticisms to this maintainability assessment formula have been reported in the literature \cite{8,9,19}. They are related to the fact that average values are used in the computation, ignoring the real distribution of values, or that the defined threshold values are not very accurate. Also, the index was defined for modular and procedural programming languages, thus not taking into consideration object oriented features that defined new relations such as inheritance, coupling and cohesion. These have been reported to have a considerable effect on maintainability \cite{7,22,21,5}.

\subsection{ARiSA Compendium Model}
\label{sec:ARISAModel}
The \textit{Compendium of Software Quality Standards and Metrics}\footnote{http://www.arisa.se/compendium/} was created by ARiSA and researchers from the Linnaeus University. It aims to study the relation between software quality characteristics and software metric values. The Compendium models software quality according to the ISO 9126, an older version in the ISO family for software quality standards. Like the more recent ISO 25010 incarnation, it is a hierarchical model made up of six characteristics, which in turn have 27 sub-characteristics. Similar with ISO 25010, \textit{Maintainability} is one of the characteristics, with sub-characteristics \textit{Analyzability}, \textit{Changeability}, \textit{Compliance}, \textit{Stability} and \textit{Testability}. For each characteristic, with the notable exception of Compliance, the set of influencing metrics is provided. For each metric influence, the Compendium details the direction and strength of the influence, as detailed using Table \ref{tab:ARISAMetrics}. The direction of the influence can be direct, or inverse, represented using upward, or downward chevrons, respectively. These illustrate whether increased values for the given metric lead to an improvement or degradation of maintainability. The number or chevrons represent the strength of this correlation, with two chevrons representing a stronger relation. For example, the weighted method count (WMC) metric relates strongly and inversely with analyzability, changeability and testability, and inversely (but not strongly) with stability. 

The VizzMaintenance \cite{11} Eclipse plugin implements a quantitative model of class-level maintainability. It is based on the relations from the Compendium and uses a number of structural, complexity and design class-level object-oriented metrics that are shown in Table \ref{tab:ARISAMetrics}. They are the coupling between objects (CBO), data abstraction coupling (DAC), depth of inheritance tree (DIT), locality of data (LD), lack of cohesion in methods (LCOM) and its improved variant (ILCOM), message pass coupling (MPC), number of children (NOC), tight class cohesion (TCC), lines of code (LOC), number of attributes and methods (NAM), number of methods (NOM), response for class (RFC), weighted method count (WMC), number of classes in cycle (CYC) length of names (LEN) and lack of documentation (LOD). They are formally defined within the Compendium \cite{11} and were used in previous research \cite{1,4}. 

The proposed quantitative model relies on the relations presented in Table \ref{tab:ARISAMetrics} and the extracted metric values. The level of maintainability is calculated for each class on a $[0,1]$ scale, with smaller values representing improved maintainability. First, the percentage of metric values within the top or bottom 15\% of each metric's value range across all classes is calculated. Then, they are aggregated across the four criteria according to the direction and strength of the relations shown in Table \ref{tab:ARISAMetrics}, resulting in the maintainability score of the class. As such, a score of 0 means that none of the metrics has an extreme value for the given class, while a value of 1 is obtained when all metric values belong in the top (or bottom) 15\%. As an example, let us consider a class having a single metric value in the top 15\%, that of the WMC. The analyzability score for WMC is $\frac{2}{33}$. The numerator is the weight WMC has for analyzability, and the denominator is the sum total of weights for that criteria (the number of chevrons). The influence of WMC in Changeability is $\frac{2}{34}$, in Stability it is $\frac{1}{26}$ and in testability it is $\frac{2}{33}$. As such, the maintainability score will be calculated as $\frac{\frac{2}{33}+\frac{2}{34}+\frac{1}{26}+\frac{2}{33}}{4}\approx0.0546$, or 5.46\%.

When compared with the MI, the ARiSA model employs a wider selection of metrics. In addition to the commonly used LOC metric, it also employs the WMC as a complexity metric, together with many well-known object-oriented ones, covering object-oriented concerns such as cohesion, coupling and inheritance. While the MI can be calculated at several granularity levels, by default the ARiSA model is limited to class level. In order to scale it to system level, we calculate its geometric mean value across all system classes.

\begin{table}[t]
\caption{Metric influences on maintainability according to the {ARiSA} Model \cite{11}}\label{tab:ARISAMetrics} \centering
\begin{tabular}
{rUUUUUUUUU||UUUUU||UUU}
  &
  \rotatebox{90}{CBO} & 
  \rotatebox{90}{DAC} & 
  \rotatebox{90}{DIT} &
  \rotatebox{90}{LD} &
  \rotatebox{90}{LCOM} &
  \rotatebox{90}{ILCOM} &
  \rotatebox{90}{MPC} &
  \rotatebox{90}{NOC} &
  \rotatebox{90}{TCC} & 
  \rotatebox{90}{LOC} &
  \rotatebox{90}{NAM} &
  \rotatebox{90}{NOM} &
  \rotatebox{90}{RFC} &
  \rotatebox{90}{WMC} & 
  \rotatebox{90}{CYC} &
  \rotatebox{90}{LEN} &
  \rotatebox{90}{LOD} \\ \cline{2-18}
  \multicolumn{1}{c|}{Analyzability} & \faAngleDoubleDown & \faAngleDoubleDown & \faAngleDoubleDown & \faAngleDoubleUp & \faAngleDoubleDown & \faAngleDoubleDown & \faAngleDoubleDown & \faAngleDown & \faAngleDoubleUp & \faAngleDoubleDown & \faAngleDoubleDown & \faAngleDoubleDown & \faAngleDoubleDown & \faAngleDoubleDown & \faAngleDoubleDown & \faAngleDoubleDown & \multicolumn{1}{c|}{\faAngleDoubleDown} \\
  \multicolumn{1}{c|}{Changeability} & \faAngleDoubleDown & \faAngleDoubleDown & \faAngleDoubleDown & \faAngleDoubleUp & \faAngleDoubleDown & \faAngleDoubleDown & \faAngleDoubleDown & \faAngleDoubleDown & \faAngleDoubleUp & \faAngleDoubleDown & \faAngleDoubleDown & \faAngleDoubleDown & \faAngleDoubleDown & \faAngleDoubleDown & \faAngleDoubleDown & \faAngleDoubleDown & \multicolumn{1}{c|}{\faAngleDoubleDown} \\
  \multicolumn{1}{c|}{Stability} & \faAngleDoubleDown & \faAngleDoubleDown & \faAngleDown & \faAngleDoubleUp & \faAngleDoubleDown & \faAngleDoubleDown & \faAngleDoubleDown & \faAngleDown & \faAngleDoubleUp & \faAngleDown & \faAngleDown & \faAngleDown & \faAngleDown & \faAngleDown & \faAngleDoubleDown & \faAngleDoubleDown & \multicolumn{1}{c|}{\faAngleDown} \\
  \multicolumn{1}{c|}{Testability} & \faAngleDoubleDown & \faAngleDoubleDown & \faAngleDoubleDown & \faAngleDoubleUp & \faAngleDoubleDown & \faAngleDoubleDown & \faAngleDoubleDown & \faAngleDown & \faAngleDoubleUp & \faAngleDoubleDown & \faAngleDoubleDown & \faAngleDoubleDown & \faAngleDoubleDown & \faAngleDoubleDown & \faAngleDoubleDown & \faAngleDoubleDown & \multicolumn{1}{c|}{\faAngleDoubleDown} \\ \cline{2-18}
  & \multicolumn{9}{c||}{Structure} & \multicolumn{5}{c||}{Complexity} & \multicolumn{3}{c}{Design} \\
\end{tabular}
\end{table}

\subsection{SQALE Model}
\label{sec:sqale}
The SQALE (Software Quality Assessment Based on Lifecycle Expectations) methodology was first introduced by J.L. Letouzey \cite{31} as a method to evaluate the quality of application source code, in an independent way from programming language or analysis tools. SQALE is tightly linked with the measurement of technical debt\footnote{Found as design debt in some sources}, especially in the context of Agile development methodologies. The first definition for technical debt was provided in 1992 \cite{29} and predates the SQALE model. Cunningham borrowed terminology from the financial sector and compared shipping immature code with \textit{"going into debt"}, and opined that doing so was fine \textit{"so long as it is paid back promptly with a rewrite"} \cite{29}. More recently, Fowler agreed that the presence of technical debt showed that delivering functionality to customers was prioritized above software quality \cite{30}. Given the focus from both researchers and practitioners on controlling software quality resulted in several tools that implement SQALE in order to produce a quantitative assessment of code quality.

Perhaps the most well-known such tool is the SonarQube platform for code quality and security. Its entry-level \textit{Community Edition} is free and open-source. Analysis support is provided through language-specific plugins, with the free version providing the required plugins for analyzing source code in 15 languages including Java, XML and HTML. Support for additional languages or features can be deployed in the form of plugins; for instance, C++ code can be analyzed using a free, community developed plugin\footnote{\url{https://github.com/SonarOpenCommunity/sonar-cxx}}. Plugins usually include a number of rules\footnote{\url{https://docs.sonarqube.org/latest/user-guide/rules/}}, against which the source code's abstract syntax tree is checked during analysis. 

Each rule is characterized by the programming language it applies to, its type, associated tags and severity. Rule type is one of \textit{maintainability} (code smell), \textit{reliability} (bug) or \textit{security} (vulnerability). Tags serve to provide a finer-grained characterization, each rule being associated with one or more tags\footnote{\url{https://docs.sonarqube.org/latest/user-guide/built-in-rule-tags/}} such as \textit{unused}, \textit{performance} or \textit{brain-overload} (e.g. when code complexity is too high). Breaking a rule results in an \textit{issue}, which inherits its characteristics from the rule that was broken. For example, Java rule \textit{S1067} states that \textit{"Expressions should not be too complex"}. It generates critical severity issues that are tagged with \textit{brain-overload} for expressions that include more than 3 operators. The time estimated to fix the issue is a 5 minute constant time to which 1 minute is added for each additional operator above the threshold. 

An application's total technical debt is calculated as the sum of the estimated times required to fix all detected issues. SonarQube normalizes the level of technical debt relevant to application size using the \textit{Technical Debt Ratio} (TDR), with $TDR = \frac{TD}{DevTime}$; $TD$ represents total technical debt quantified in minutes, while $DevTime$ represents the total time required to develop the system, with 30 minutes of time required to develop 1 line of production level code. The application is graded according to the SQALE rating between A (best value, $TDR < 5\%$) and E (worst value, $TDR \geq 50\%$). SQALE provides a high-level, evidence-backed and easy to understand interpretation of the system's internal quality. In our case study we calculate SQALE ratings using SonarQube version 8.2, which integrates the Eclipse Java compiler and uses more than 550 rules to detect potential issues in source code.

While SonarQube and similar tools provide quantitative models of software quality, existing research also pointed out some existing pitfalls. Authors of a large-scale case study \cite{41} showed that many of the reported issues remained unfixed, which could be the result of these tools reporting many false-positive, or low-importance results. A study of SonarQube's default rules \cite{42} also showed most of them having limited fault-proneness. These findings are also mirrored in our work \cite{43}, where we've shown that issue lifetimes are not correlated with severity or associated tags.    

\section{State of the Art}
\label{sec:RelatedWork}
The role and impact of maintainability as a software quality factor was investigated in existing literature \cite{27,8,24,37}. The SIG Maintainability model \cite{27,8} is based on the idea of relating different source code properties such as volume, complexity and unit testing with the sub-characteristics of maintainability as described according to the ISO 9126 model \cite{35}. The SIG Maintainability model was evaluated on a large number of software applications. Authors of \cite{24} proposed a framework in which quality characteristics defined according to the ISO 25010 model \cite{23} could be assessed directly or indirectly by associated measures that can be easily computed using existing software tooling. The framework remains a proof of concept with more measures required for consideration before the measurement of quality characteristics such as maintainability becomes possible.

The ARiSA model \cite{33,12} detailed in Section \ref{sec:ARISAModel} remains one of the most exhaustive studies that analyzes the relations between a significant number of software metrics and quality factors and sub-factors as defined according to ISO 9126.

The influence object-oriented metrics have on maintainability received a continuous interest in the research community ever since their introduction \cite{7}, with existing research showing the existence of a relation between maintainability, coupling and cohesion \cite{38,39,40,22,5}. The influence different metrics have on maintainability has also received intense scrutiny. However, we find that in many cases author conclusions are limited to the identified relation between a singular metric and target system maintainability \cite{12,22}. While important in itself, these do not provide a definitive quantitative model for maintainability, partly due to their strong empirical nature. Thus, in order to develop more precise and easily applicable methods for assessing software quality characteristics we find that more investigations need to be carried out and reported. 

As such, the distinctive feature of our study is that it describes and analyzes three approaches of different complexity that enable an evaluation of maintainability in the case of large applications. Furthermore, we analyze and compare results across the application versions and maintainability models themselves in order to improve our understanding of the evolution of open-source applications on one hand, as well as the applicability, strengths and weaknesses of maintainability models on the other.

\section{Case Study}
\label{sec:CaseStudy}
The presented case study is the direct continuation of the work presented in \cite{44}. The work was organized and carried out according to currently defined best practices \cite{17,45}. We started by stating the main objective of our work, which we distilled into four research questions. We structured the current section according to Höst and Runeson's methodology \cite{17}. We first discuss the selection of target applications, after which we present the data collection process. We used Section \ref{sec:analysis} to discuss the results of our analysis, after which we address the threats to our study's validity.

\subsection{Research Questions}
\label{sec:rq}
We defined our work's main objective using the goal-question-metric approach \cite{18} to be \textit{"study the maintainability of evolving open-source software using quantitative software quality models"}. We refined our stated objectives into four research questions. They serve to guide the analysis phase of the study, as well as to provide an in-depth view when compared with our previous work \cite{44}.

\bm{$RQ_{1}$}: \textit{What is the correlation between application size and maintainability?} In our previous work \cite{44} we have disproved the na\"ive expectation that lower maintainability is reported for larger applications. However, we employ $RQ_{1}$ to ensure that maintainability scores reported using the proposed quantitative models are not excessively influenced by software size. While in our previous work \cite{44} we have examined this relation using the number of classes as a proxy for system size, we extend our investigation to cover the number of packages, methods and lines of code. We aim to employ system size measurements in order to study their effect on reported maintainability, as both the MI and ARiSA models include class and line counts in their assessment.

\bm{$RQ_{2}$}: \textit{What drives maintainability changes between application versions?} In our previous study we identified important changes in the maintainability scores reported for the target applications. We expect the answer to $RQ_{2}$ will help us identify the rationale behind the large changes in maintainability reported in each of the target applications studied in our previous work. We aim to triangulate collected data \cite{17} by carrying out a cross-application examination. We expect this will facilitate identifying common causes and help alleviate external threats to our study. In order to properly contextualize observed changes to maintainability, we carry out a detailed manual source code examination.

\bm{$RQ_{3}$}: \textit{How are maintainability changes reflected at the package level?} We employ $RQ_{1}$ in order to study the relation between reported maintainability and software size, each measured according to several quantitative metrics. Then, the answer to $RQ_{2}$ helps determine the amplitude and rationale behind the reported changes. We take the following step via $RQ_{3}$, where we carry out a finer grained analysis at package level, in order to improve our understanding of the impact software evolution has on application component maintainability.

\bm{$RQ_{4}$}: \textit{What are the strengths and weaknesses of the proposed maintainability models?} In our previous research we determined the $TDR$ to be the most relevant quantitative model from a software development perspective \cite{44}. However, we also discovered that both the ARiSA model and the MI can provide actionable information in the right context. This is especially true since the ARiSA model was created for class-level usage, while the MI works from system down to method levels. As such, as part of our data analysis we examine our answers to $RQ_{2}$ and $RQ_{3}$ and highlight the insight  that each model can provide together with its drawbacks.

\begin{table}[t]
\caption{Information about the earliest and latest target application versions in our study}\label{tab:TargetAppVersions} \centering
    \begin{tabular}{ccccccc}
    \multirow{2}{*}{Application} & \multirow{2}{*}{Version} & \multirow{2}{*}{Release Date} & \multirow{2}{*}{Statements} & \multicolumn{3}{c}{Maintainability Rating} \\
    &  &  & & MI & ARiSA & SQALE \\ \hline
    \multirow{2}{*}{FreeMind} & 0.0.3 & July 9, 2000 & 1,359 & 81.30 & 0.22 & 3.10 \\
    & 1.1.0Beta2 & Feb 5, 2016 & 20,133 & 73.92 & 0.19 & 3.2 \\ \hline
    \multirow{2}{*}{jEdit} & 2.3pre2 & Jan 29, 2000 & 12,150 & 73.38 & 0.16 & 3.00 \\
    & 5.5.0 & April 9, 2018 & 43,875 & 66.90 & 0.17 & 3.50 \\ \hline
    \multirow{2}{*}{TuxGuitar} & 0.1pre & June 18, 2006 & 4,863 & 71.53 & 0.12 & 2.10 \\
    & 1.5.3 & Dec 10, 2019 & 51,589 & 75.99 & 0.17 & 1.20 \\ \hline
    \end{tabular}
\end{table}

\subsection{Target Applications}
\label{sec:targetapp}
Since the present paper builds upon and expands our previous research \cite{44}, we maintained our selection of target applications. In this section we reiterate our rationale and briefly discuss inclusion criteria. Our main goal was to select a number of complex, widely-used and open-source applications that facilitate evaluating maintainability in the context of software evolution. Previous empirical research in open-source software has shown that many of these systems go through development hiatuses, or are abandoned by the original developers \cite{46}. In other cases, available source code is incomplete, with missing modules or libraries, or contains compile errors \cite{1}. Other applications include complex dependencies which are required to compile or run them, such as Internet services, database servers or the presence of additional equipment.

Taking these considerations into account, we set our inclusion criteria to applications with a long development history and no external dependencies. In order to allow comparing results across applications and in order to alleviate external threats to our study, we limited ourselves to GUI-driven applications developed using the Java platform.

\begin{figure*}[!ht]
    \captionsetup[subfigure]{labelformat=empty,justification=centering}
    \centering
    \begin{subfigure}[b]{\textwidth}
        \includegraphics[width=\textwidth]{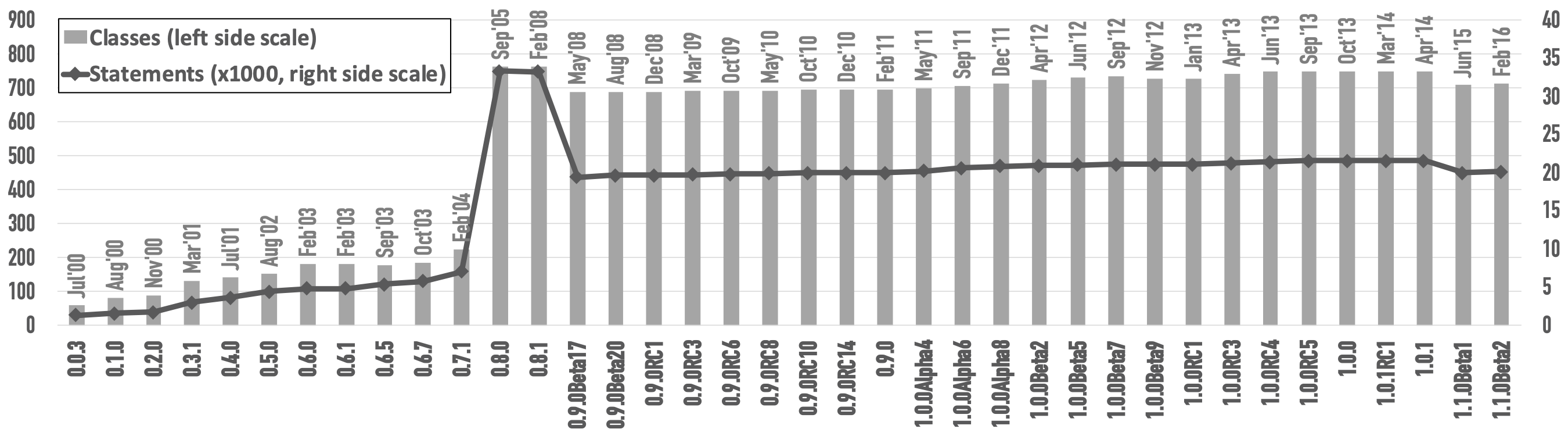}
    \end{subfigure}
    
    \begin{subfigure}[b]{\textwidth}
        \includegraphics[width=\textwidth]{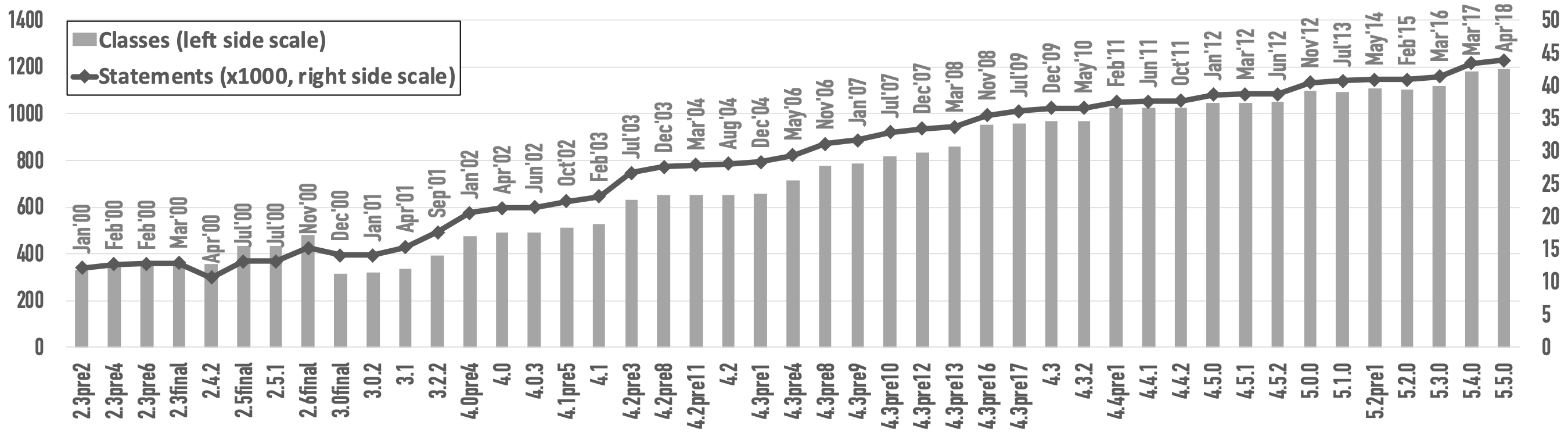}
    \end{subfigure}
    \par\medskip
    \begin{subfigure}[b]{\textwidth}
        \includegraphics[width=\textwidth]{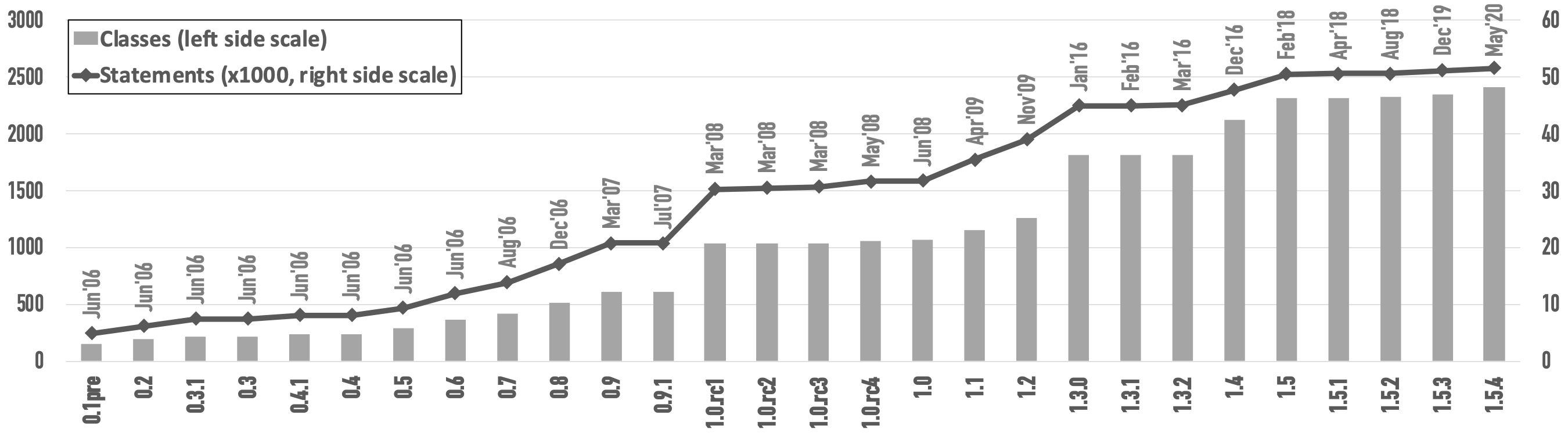}
    \end{subfigure}
    \caption{Size and release date information for FreeMind (top), jEdit (middle) and TuxGuitar (bottom) versions included in our study.}
    \label{fig:ClassLOC}
\end{figure*}

Our selection process resulted in three Java applications. Each of them is available under permissive open-source licensing, has a fully documented development history including an important number of version releases and a consistent user base. They are the FreeMind\footnote{\url{http://freemind.sourceforge.net/wiki/index.php/Main_Page}} mind mapper, the jEdit\footnote{\url{http://jedit.org}} text editor and the TuxGuitar\footnote{\url{http://www.tuxguitar.com.ar}} tablature editor. Table \ref{tab:TargetAppVersions} provides relevant information for the first and last target application version included in our case study. We refer to all releases included in our study using the version numbers assigned to them by developers. We believe this provides additional context on the magnitude of expected changes between versions, and facilitates replicating our results, as it allows third parties to unambiguously identify them within the code base. Furthermore, as we identified several hiatuses in the development of the studied applications, we found version numbers were more representative than release dates.  

\textbf{FreeMind} is a mind-mapping application with a consistent user base, rich functionalities and support for plugin development and integration. The first version in our study is 0.0.3. Released in July, 2000, it consisted of around 1,350 code statements and around 60 classes, which make it the smallest release in our study. This is reflected on a functional level, with early versions of FreeMind having limited functionalities, in contrast with later versions. We take this into account in our research when studying the difference between early and mature application versions. Several of its versions were used in previous empirical research \cite{2}. Figure \ref{fig:ClassLOC} illustrates the development across the versions in our case study using system size and release dates. Versions 0.8.0 and 0.8.1 show an important increase in system size, which is tempered in version 0.9.0Beta17, after which system size remains stable. We also note the $2\frac{1}{2}$ years of hiatus between versions 0.8.0 and 0.8.1. Major changes are recorded for version 0.9.0Beta17, released only 3 months after the previous one. While the most recent version released at the start of 2016, FreeMind maintained a consistent user base, with 681k application downloads during last year and over 21 million over its lifetime\footnote{Download data points from \url{https://sourceforge.net/}, only consider application releases. Recorded August $25^{th}$, 2020}.

\textbf{jEdit} is a plugin-able text-editor targeted towards programmers. As shown in Figure \ref{fig:ClassLOC}, its first public version, 2.3pre2 was released in January 2000. Having over 300 classes and 12,000 statements, it is the most polished entry version in our study. In opposition to FreeMind, we did not record multi-year hiatuses during the development of jEdit. Class and statement counts showed a gradual, but steady increase version to version, which we found reflected at the user experience and functional levels. jEdit was also the subject of software engineering research that targeted GUI testing \cite{2,3} and software quality \cite{4,5}. The application has managed to maintain a large user base, having over 92k application downloads last year and over 5.8 million over its lifetime.

\textbf{TuxGuitar} is a tablature editor with multi-track support, which support data import and export across multiple formats. This is implemented in the form of plugins that are included in the default code distribution, which we included in our case study. TuxGuitar was developed with support for several GUI toolkits, and we selected to use its SWT implementation across all versions. As illustrated in Figure \ref{fig:ClassLOC}, TuxGuitar's evolution is similar to that of jEdit, with a steady increase in application size across most versions. While its development seemed to be halted between versions 1.2 and 1.3.0, its latest version was released in 2020, with the project being actively developed. TuxGuitar also has a consistent user base, recording 266k application downloads during the last year and over 6.9 million over its lifetime. 

\subsection{Data Collection}
We limited our selection to publicly released versions in order to address the risk of compiler errors or missing libraries, as reported by previous research \cite{1}. We handled the case of many incremental version releases in the span of days by only considering the last of them, which helps keep the number of versions manageable. This resulted in 38 releases of FreeMind, 45 releases of jEdit and 28 releases of TuxGuitar included in our study. Each release was then imported into an IDE, where a manual examination of its source code was carried out. A common recurring issue concerned the presence of library code shipped together with application source code. Several jEdit versions included code for the \textit{com.microstar.xml} parser or \textit{BeanShell} interpreter. Our solution was to extract these into a separate library that was added to the application classpath. FreeMind and jEdit source code was analyzed without any plugin code, although both applications provide plugin support. In the case of TuxGuitar, we kept the data import/export plugins included in the official distribution.   

Metric data was extracted using the VizzMaintenance plugin for Eclipse, which was also used to calculate the ARiSA maintainability score of each class. We employed the Metrics Reloaded plugin for IntelliJ to calculate the components of the MI, while the SQALE rating was obtained using the community edition of SonarQube 8.2. 

\subsection{Analysis}
\label{sec:analysis}
In this section we present the most important results of our analysis, structured according to the research questions defined in Section \ref{sec:rq}. In order to facilitate replicating or extending our results, we made available the entire set of collected and processed metric data \cite{55}.

\begin{table}[t]
\caption{Spearman rank correlation between software size according to package, class, method or statement count and system maintainability. FreeMind data on top row, jEdit on middle row and TuxGuitar on bottom row.}\label{tab:MaintainabilitySizeCorrelation} \centering
\begin{tabular}{rrrrr}
     & Package & { }Class & { }Method & { }Statement  \\ \cline{2-5}
  \multirow{3}{*}{MI} & -0.46 & -0.67 & -0.71 & -0.77 \\
                      & -0.32 &	-0.29 &	-0.40 & -0.45 \\
                      & 0.76 & 0.71 & 0.69 & 0.57 \\ \cline{1-5}
  \multirow{3}{*}{ARiSA} & -0.29 & -0.49 & -0.53 & -0.62 \\
                         & -0.51 & -0.53 & -0.54 & -0.55 \\
                         & 0.64	& 0.63 & 0.64 & 0.69 \\ \cline{1-5}
  \multirow{3}{*}{SQALE} & -0.23 & 0.10 & 0.19 & 0.38 \\
                         & 0.12 & 0.14 & 0.21 & 0.24 \\
                         & -0.68 & -0.73 & -0.74 & -0.79 \\ \cline{1-5}
\end{tabular}
\end{table}

\subsubsection{\bm{$RQ_{1}$}: \textit{What is the correlation between application size and maintainability?}} In our previous research \cite{5,44} we showed that maintainability measured according to quantitative metrics was not correlated with software size, at least not when the latter was expressed using the number of the system's classes. We extended our investigation to also cover the number of a system's packages, methods and statements\footnote{Collected using the Metrics Reloaded plugin for IntelliJ}. Since target applications were developed using Java, there was a strong and expected correlation between class and source file counts, so this evaluation was omitted.

We first carried out a Spearman rank correlation between the size measures for each application. We found very high correlation between all measures for application size ($\rho \geq 0.8$), especially between the number of classes, methods and statements ($\rho \geq 0.96$) for each target application. 

We repeated the correlation analysis between the size measurements and reported values for maintainability, which we report using Table \ref{tab:MaintainabilitySizeCorrelation}. Note that higher scores correspond to a decrease in maintainability according to the ARiSA and SQALE models, and an increase according to the MI. Results for FreeMind and jEdit show similarity across all three models. We note that increased values for the MI are accounted for by joint increases across its components (statement count, Halstead volume and cyclomatic complexity). Values produced by our generalization of the ARiSA model are skewed by small files added in later software versions; these keep mean values low, leading to what we believe are under-reported changes to maintainability. The SQALE model is driven by static analysis of the abstract syntax tree, and is not directly influenced by size-related metric values. This also explains the weak correlation with the number of statements, which is the lowest-level size metric considered.

\begin{figure*}[t]
    \captionsetup[subfigure]{labelformat=empty,justification=centering}
    \centering
    \begin{subfigure}[b]{\textwidth}
        \includegraphics[width=\textwidth]{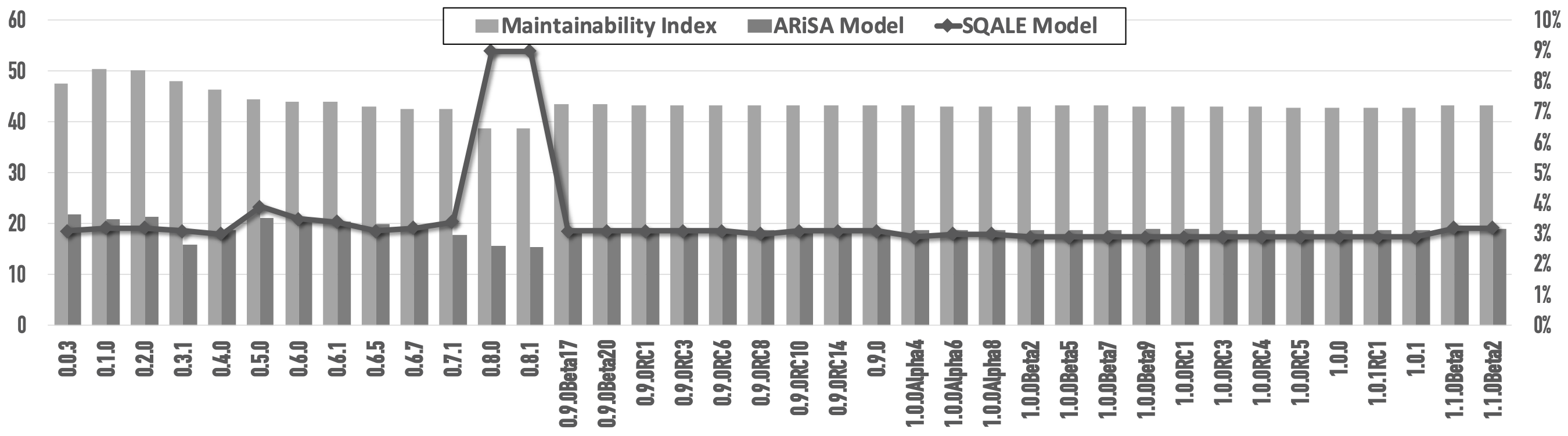}
    \end{subfigure}
    
    \begin{subfigure}[b]{\textwidth}
        \includegraphics[width=\textwidth]{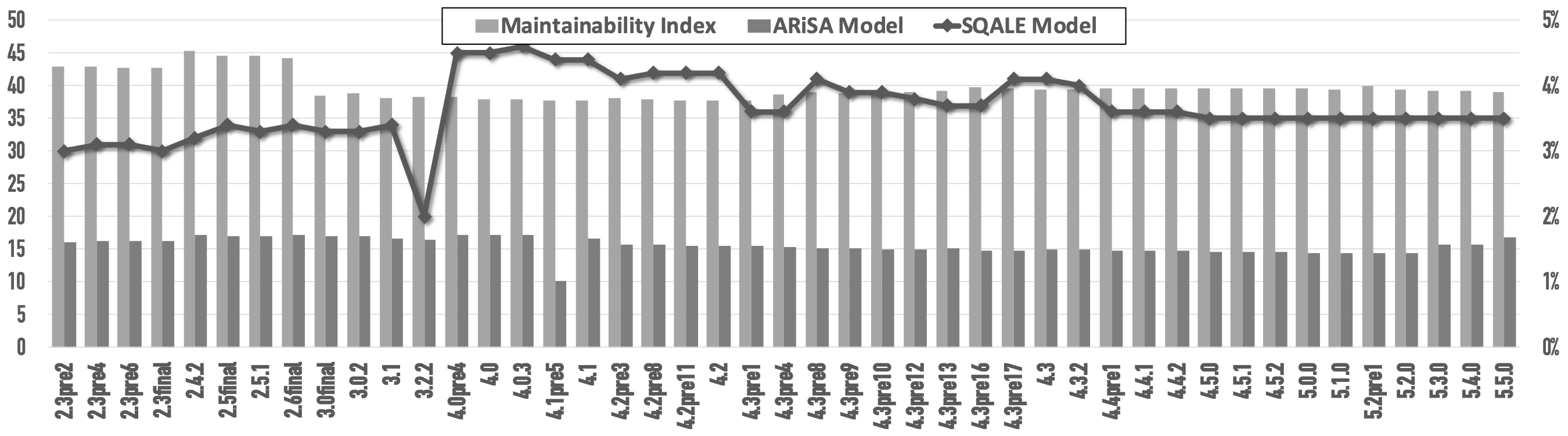}
    \end{subfigure}
    \par\medskip
    \begin{subfigure}[b]{\textwidth}
        \includegraphics[width=\textwidth]{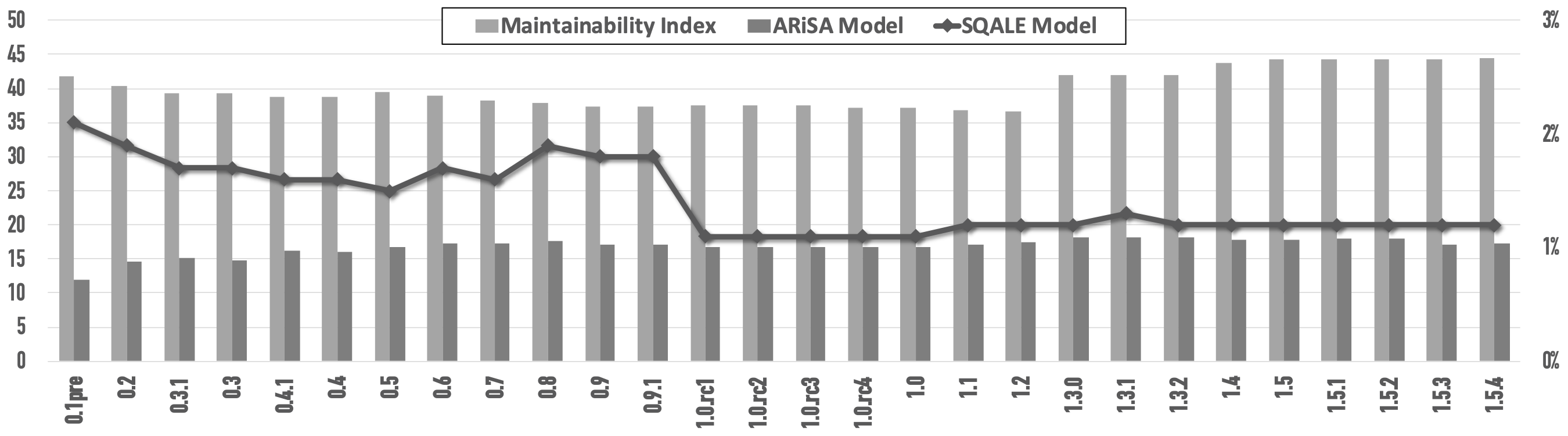}
    \end{subfigure}
    \caption{Maintainability of FreeMind (top), jEdit (middle) and TuxGuitar (bottom) versions in our study. SQALE model uses the scale on the right side.}
    \label{fig:maintainability}
\end{figure*}

As shown in Figure \ref{fig:maintainability}, TuxGuitar was evaluated as having very good maintainability \cite{44}. All releases remained well below the $5\%$ threshold required to receive an $A$ rating according to SQALE. We believe this to be the result of a conscientious effort on the behalf of its developers. Important increases to system size, such as those for versions 1.0rc1, 1.3 and 1.5 did not have an important effect on measured maintainability. In version 1.0rc1, an increase in system size was actually coupled with improved maintainability according to SQALE.

Our analysis showed the MI and ARiSA models to be influenced by software size, which is known to have a confounding effect \cite{47}. The SQALE model did not appear to have been influenced by it, as it does not rely on size-related software metrics.

\subsubsection{\bm{$RQ_{2}$}: \textit{What drives maintainability changes between application versions?}}
We base our answer to $RQ_{2}$ on the data from Figure \ref{fig:maintainability}, which shows system-level maintainability according to the three models. Data is normalized to the [0, 100] range. Our previous research \cite{44} showed that of the proposed quantitative models, technical debt was the most suitable for evaluating system-level quality. As such, we focus on the SQALE rating to quantify system-level maintainability \cite{44}. According to it, most application versions have good maintainability, with most studied versions receiving an $A$ rating; the only exceptions were FreeMind versions 0.8.0 and 0.8.1, which earned a $B$ SQALE rating. We also note that most TuxGuitar versions have a $TDR \leq 2\%$, as during our previous evaluation we found evidence of concerted developer action to improve software quality. 

Examining the data in Figure \ref{fig:maintainability} revealed that system-level maintainability did not suffer major changes across most versions. As such, we identified key versions \cite{44} during which quantitative changes were detected. In the case of FreeMind, versions 0.8.* were the result of significant application development that increased application size from 12.5k LOC to 65.5k LOC, with an additional 370 days worth of technical debt added \cite{43,44}. Most of the added debt was fixed in version 0.9.0Beta17 with no loss to functionality; our detailed analysis of subsequent versions only revealed small-scale maintainability changes \cite{43}.

Our evaluation of jEdit version 4.0pre4 revealed that additional functionalities such as improved management of the text area, buffer events and the document model, implemented using 11k LOC added an extra month of technical debt. However, our detailed examination \cite{43} revealed that versions after 4.0 gradually reduced the level of debt and the addition of significant additional quality issues appears to have been avoided.

Most of the changes observed within TuxGuitar were of smaller significance, as it already presented very good maintainability. The most significant version shown in Figure \ref{fig:maintainability} is 1.0rc1; here, we observed that extensive refactoring efforts on existing debt were coupled with the introduction of additional issues \cite{43}, most likely as part of the additional support for the song collection browser and the inclusion of new plugins. Overall, technical debt was improved to a level that was maintained until the most recent released version.

\subsubsection{\bm{$RQ_{3}$}: \textit{How are maintainability changes reflected at the package level?}}
Our previous evaluation \cite{44} revealed that among studied models, SQALE was the one best suited for system and package-level quality assessment. As such, we used SonarQube's estimation of required maintainability effort at package level for each of the application versions in our study. We identified core packages that existed within all studied versions, as well as packages introduced at some point and removed at a later time. We found this to be typical of TuxGuitar, for which we counted a total of 354 packages across all versions. On the other hand, jEdit's entire code base consisted of 30 packages, while the FreeMind code base covering all releases was comprised of 41 packages.

\begin{figure*}[t]
    \captionsetup[subfigure]{labelformat=empty,justification=centering}
    \centering
    \begin{subfigure}[b]{\textwidth}
        \includegraphics[width=\textwidth]{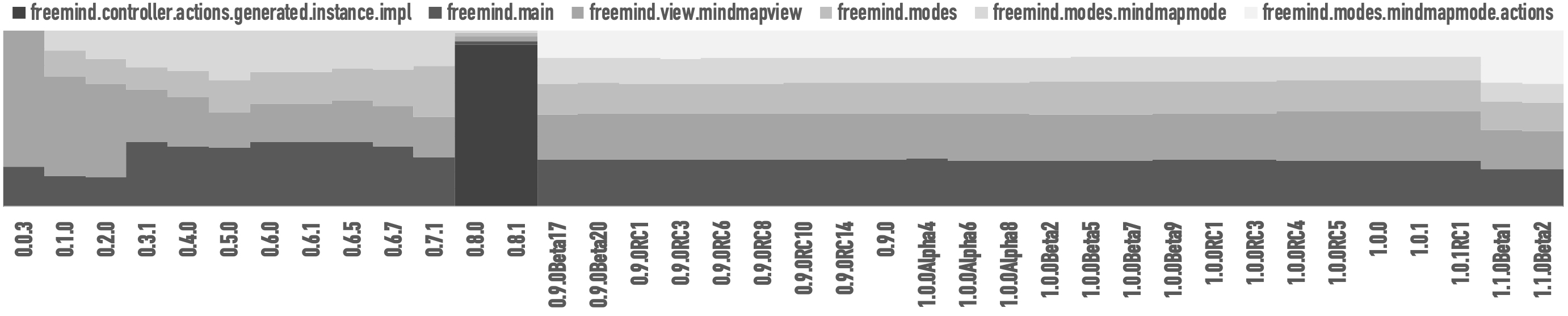}
    \end{subfigure}
    
    \begin{subfigure}[b]{\textwidth}
        \includegraphics[width=\textwidth]{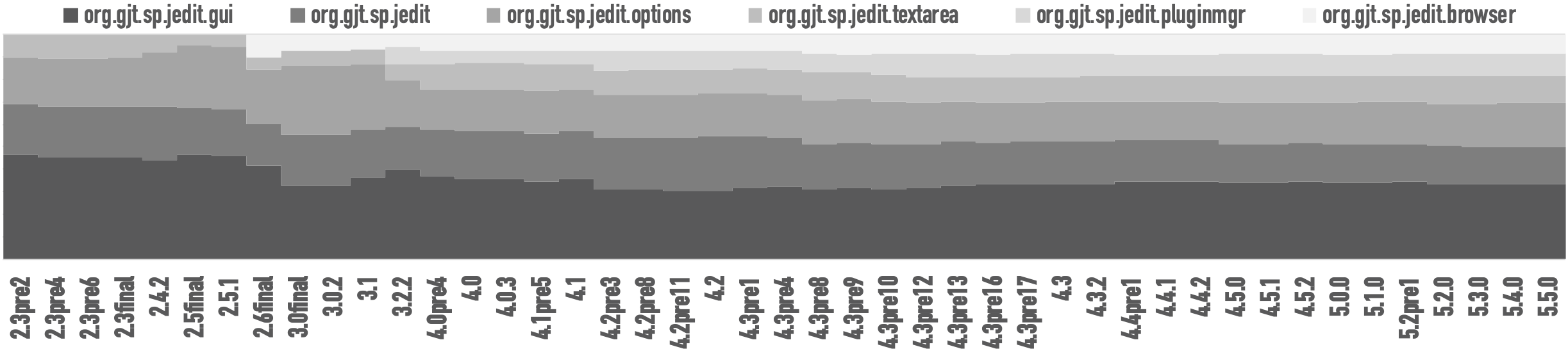}
    \end{subfigure}
    \par\medskip
    \begin{subfigure}[b]{\textwidth}
        \includegraphics[width=\textwidth]{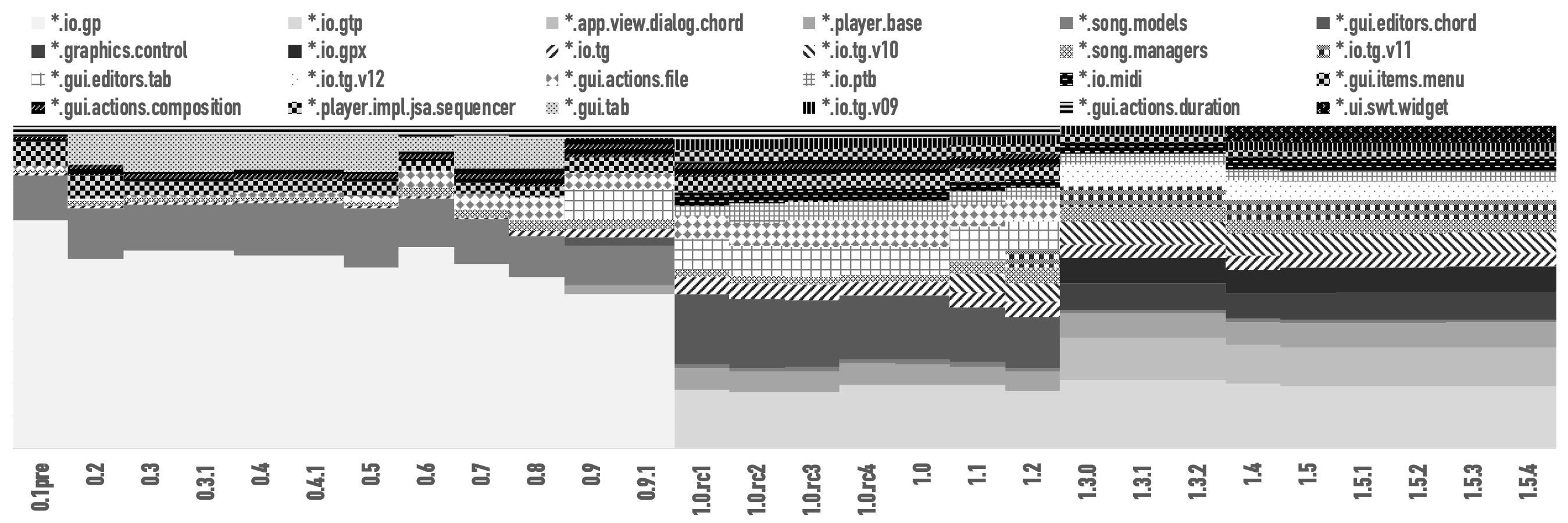}
    \end{subfigure}
    \caption{Estimated maintainability effort at package level according to the SQALE model for FreeMind (top), jEdit (middle) and TuxGuitar (bottom, * stands for \textit{org.herac.tuxguitar}) versions. Represented packages account for at least half of total effort per application.}
    \label{fig:package_td}
\end{figure*}

We represent the estimated time to address at least half the total maintainability effort for each application in Figure \ref{fig:package_td}. We show that most maintainability issues were concentrated on a small subset of application packages. For instance, the six packages represented for jEdit account for almost 80\% total maintenance effort, while the 24 packages illustrated for TuxGuitar cover half the required effort. 

Figure \ref{fig:package_td} shows FreeMind and jEdit versions to be very stable with regards to the distribution of the required maintenance effort. In the case of FreeMind, we discovered it was mainly generated code from the \textit{[...].generated.instance.impl} package that caused the severe decrease in maintainability, while in the remaining application packages the maintenance effort did not change significantly. For TuxGuitar, we noted the changes in versions 1.0rc1 and 1.3.0. While our previous evaluation already showed system maintainability to be affected within these versions \cite{44}, it was package-level examination that revealed changes to plugin code from the \textit{org.herac.tuxguitar.io.gp} package as the cause of changes in version 1.0rc1. 

Figure \ref{fig:package_td} also reveals information about application architecture. Both FreeMind and jEdit were built around a relatively small, but constant set of packages and suffered most changes during the development of their early versions \cite{5}. In the case of TuxGuitar, we found each plugin to have a separate package, with many input-output plugins maintaining separate packages for each implementation versions. This resulted in a more complex and change-prone hierarchy.

We found that the main advantage of drilling down to package level regarded the precise identification of the locations and dimensions of the maintenance effort. When combined with a longitudinal analysis \cite{43}, a package-level evaluation can help with program comprehension and testing, as it can be used to discard application areas that have not undergone changes.  

\begin{figure*}
    \captionsetup[subfigure]{labelformat=empty,justification=centering}
    \centering
    \begin{subfigure}[b]{\textwidth}
        \includegraphics[width=\textwidth]{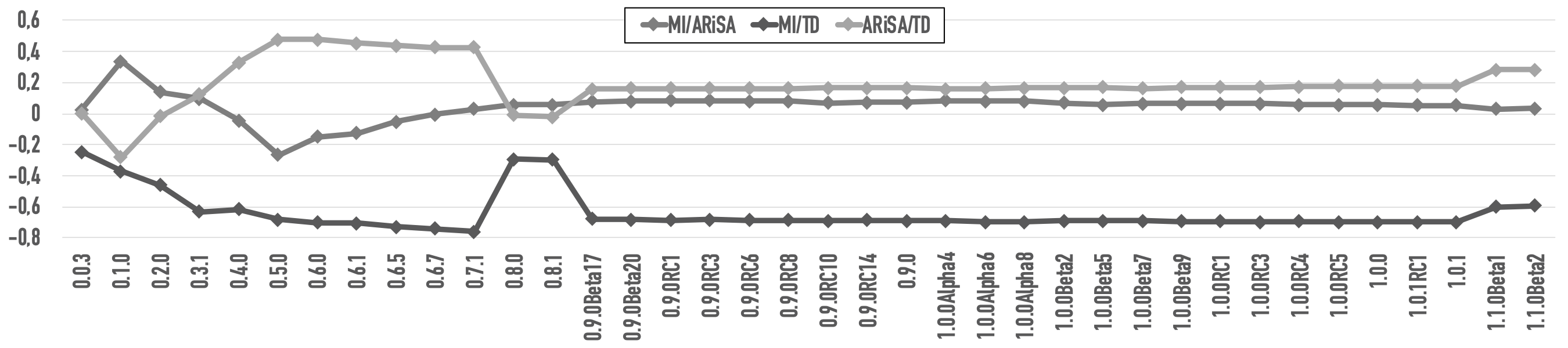}
    \end{subfigure}
    
    \begin{subfigure}[b]{\textwidth}
        \includegraphics[width=\textwidth]{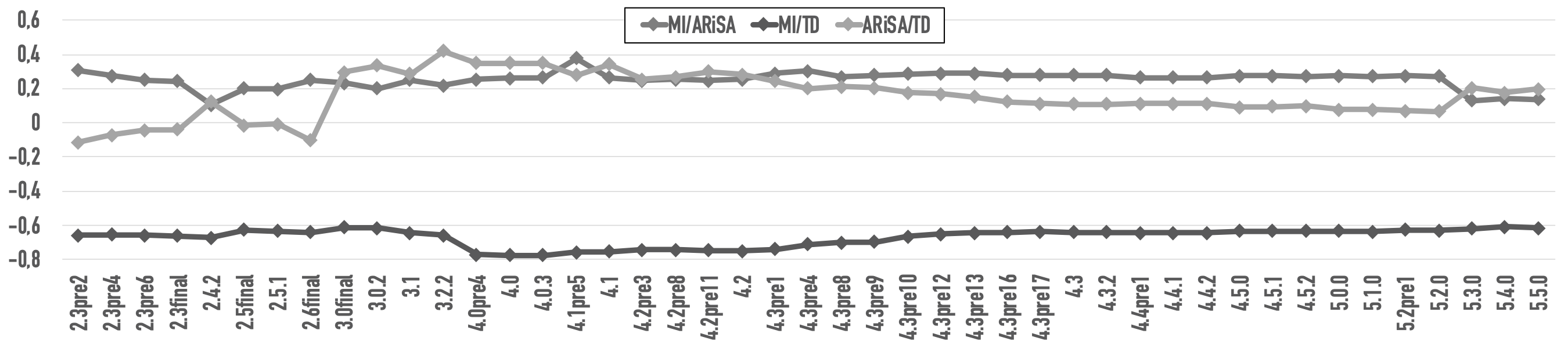}
    \end{subfigure}
    \par\medskip
    \begin{subfigure}[b]{\textwidth}
        \includegraphics[width=\textwidth]{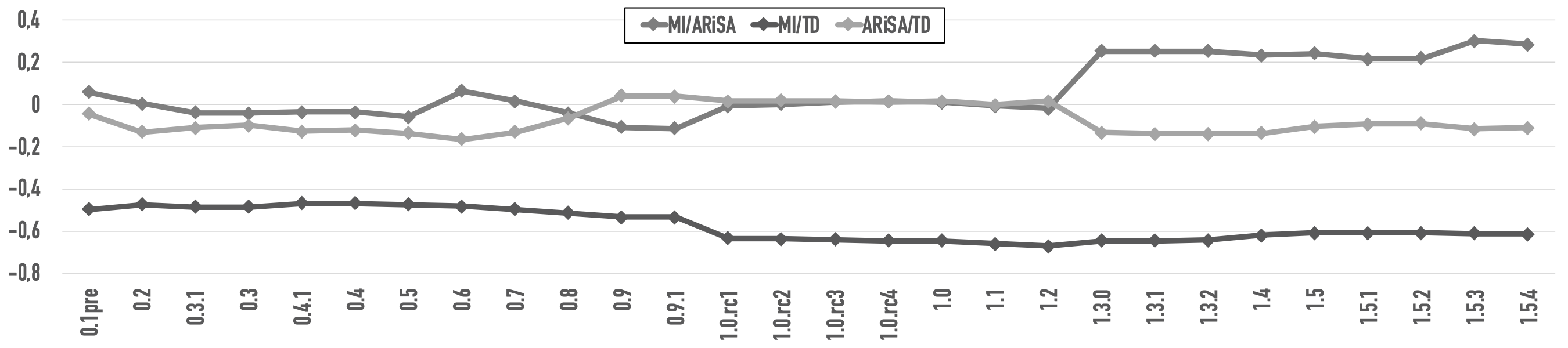}
    \end{subfigure}
    \caption{Value of the Spearman correlation coefficient between maintainability models applied at class level for FreeMind (top), jEdit (middle) and TuxGuitar (bottom) versions.}
    \label{fig:correlations}
\end{figure*}

\subsubsection{\bm{$RQ_{4}$}: \textit{What are the strengths and weaknesses of the proposed maintainability models?}}
Our answer for $RQ_{4}$ takes into account our existing research in software maintainability \cite{5,44} and the long-term evolution of technical debt \cite{43}, as well as the results of the detailed examination carried out for $RQ_{2}$ and $RQ_{3}$. We found that the SQALE model, and its implementation in the form of technical debt to be the most accurate quantitative quality measure among those studied. Technical debt evaluation provides a general assessment of a given system, but can also be employed at a finer level of granularity to uncover the root causes of detected issues \cite{43}. However, existing criticism outlined in Section \ref{sec:sqale} point against using it without prejudice to evaluate software quality.

The ARiSA model is strictly based on the evaluation of extreme values for class level object-oriented metrics. We found that aggregating class-level scores did not produce useful results, as in many cases quality issues were masked, or completely countered by large numbers of small, low complexity classes that influenced mean values. The same criticism can be brought against the MI, computing which is limited to three metrics, none of which specific to the object-oriented domain. The important advantage of the MI is that it remains language-independent and has a straightforward implementation. As such, in order to examine its potential for usage at a finer-grained level, we carried out a Spearman rank correlation between the result produced using the proposed models at class level, as shown in Figure \ref{fig:correlations}. The only consistent correlation observed occurred between the MI and technical debt with $\rho \approx -0.6$. We believe this is an indication that the MI can be employed to quickly discover code complexity issues at method and class levels. However, a more detailed examination is required in order to fully describe and characterize this result. 

We believe the ARiSA model remains well suited to discovering quality hot spots within the context of a singular application \cite{44}. While it employs an important number of object-oriented measurements, the model derives value thresholds from the evaluated system's context, making it unsuitable for cross-application and cross-version comparisons.

\subsection{Threats to Validity}
\label{sec:threats}
We structured the case study according to existing best practices \cite{17}. First, the main objective and research question were defined, after which target application selection took place. This was followed by the data collection and analysis phases. We carried out a manual examination of source code in order to complement the results from the quantitative models, and open-sourced the data to facilitate replicating or extending our study \cite{55}.

Internal threats were addressed by complementing automated evaluation with a manual examination of the source code; this step was also assisted by source code in order to prevent any observer bias. Data analysis was carried out using previously used research tooling \cite{4,5,43,44} to avoid the possibility of software defects influencing evaluation result.

Quantitative models employed were selected according with their previous use in research and practice, as well as varying implementation complexity. MI values were calculated using both statement count and LOC, while for the ARiSA model we studied the effect of calculating the final value using all three Pythagorean means. 

We addressed external threats by limiting target application selection to GUI-driven Java applications. While this limited the applicability of our study's conclusions, it also enabled data triangulation and directly comparing results across applications. To the best of our knowledge, there were no overlaps between target application development teams. Furthermore, neither of the present study's authors were involved with their development. 

The entire data set of extracted metric values together with versions processed by our tooling are open-sourced and freely available. We believe this to be the most important step required in order to solidify our results and encourage further work to extend them.

\section{Conclusion and Future Work}
\label{sec:Conclusion}
In the present paper we continued our empirical research targeting the relation between metric values and software product quality \cite{4,5,43,44}. We confirmed our initial findings regarding the independence of maintainability effort from software size \cite{44}. We also confirmed initial expectations regarding the gradual, but sustained increase in application size during development. However, we could also identify key versions where extensive refactoring kept application size and complexity in check. Another interesting observation was that mature application versions no longer introduced significant quality issues. We first observed this when studying software metric values \cite{4,54} and confirmed it through evaluating the data summarized in Figure \ref{fig:maintainability}. We believe this can be explained through an already matured application architecture, together with the existence of a core of experienced contributors. 

Our evaluation also uncovered the existence of milestone versions, characterized by significant changes at source code level and the addition of many new features. Versions such as FreeMind 0.8.0, jEdit 4.0pre4 or TuxGuitar 1.0rc1 are such examples, where changes to the application had an important effect on software quality. The case of TuxGuitar 1.0rc1 is especially worth mention, as a development milestone was coupled with refactoring efforts that lowered maintenance effort. With regards to the root causes of changes to maintainability, we consistently found the main drivers to be significant changes to application presentation, functionality and extensive refactoring.

Most of the existing research is limited to evaluating software quality at system level. In our study, we carried out a finer grained analysis at application package level in order to improve our understanding of the distribution and evolution of the maintenance effort. Figure \ref{fig:package_td} illustrates this for the most maintenance-heavy application packages. This allowed us to discover the root cause of the maintenance spike in FreeMind 0.8.0, as well as the effects of the plugin-centered architecture on the distribution and evolution of maintenance effort for TuxGuitar. Especially in the case of TuxGuitar, Figure \ref{fig:package_td} illustrates how maintenance effort was redistributed across the packages in versions with significant changes to source code.

Finally, our study provided an opportunity to examine the maintainability models themselves. We found the MI to remain useful at a very fine granularity level, and can be used at method or class level to ensure code complexity remains in check. We found the ARiSA model to be useful at application level, but its particularities preclude it from being useful when comparing applications. This can be achieved using the SQALE methodology and its implementations, which provide a language agnostic measurement scale.

Further directions targeting this research topic include extending the evaluation from two perspectives. First, to consider other types of software systems, such as mobile or distributed applications. Second, to investigate the effect the development platform and programming language have on maintenance effort. 

%
%
\bibliographystyle{splncs04}
\bibliography{references}

\end{document}